\documentclass[12pt]{article}%
\usepackage{amsfonts}
\usepackage{amsmath}
\usepackage{amssymb}
\usepackage{graphicx}
\usepackage{setspace}%
\newtheorem{theorem}{Theorem}

\newtheorem{conjecture}{Conjecture}
\newtheorem{corollary}[theorem]{Corollary}

\newtheorem{definition}{Definition}

\newtheorem{lemma}[theorem]{Lemma}

\newtheorem{problem}{Problem}

\begin{document}

\title{Quantum Mechanics as a Theory of Probability}
\author{Itamar Pitowsky\\Department of Philosophy, The Hebrew University\\e-mail: Itamarp@vms.huji.ac.il\\web site: http://edelstein.huji.ac.il/staff/pitowsky/}
\maketitle

\begin{abstract}
We develop and defend the thesis that the Hilbert space formalism of quantum
mechanics is a new theory of probability. The theory, like its classical
counterpart, consists of an algebra of events, and the probability measures
defined on it. The construction proceeds in the following steps: (a) Axioms
for the algebra of events are introduced following Birkhoff and von Neumann.
All axioms, except the one that expresses the uncertainty principle, are
shared with the classical event space. The only models for the set of axioms
are lattices of subspaces of inner product spaces over a field $K$. (b)
Another axiom due to Sol\`{e}r forces $K$ to be the field of real, or complex
numbers, or the quaternions. We suggest a probabilistic reading of Sol\`{e}r's
axiom. (c) Gleason's theorem fully characterizes the probability measures on
the algebra of events, so that Born's rule is derived. (d) Gleason's theorem
is equivalent to the existence of a certain finite set of rays, with a
particular orthogonality graph (Wondergraph). Consequently, all aspects of
quantum probability can be derived from rational probability assignments to
finite "quantum gambles". (e) All experimental aspects of entanglement- the
violation of Bell's inequality in particular- are explained as natural
outcomes of the probabilistic structure. (f) We hypothesize that even in the
absence of decoherence macroscopic entanglement can very rarely be observed,
and provide a precise conjecture to that effect. We also discuss the relation
of the present approach to quantum logic, realism and truth, and the
measurement problem.

\end{abstract}
\tableofcontents

\section{Introduction}

Discussions of the foundations of quantum mechanics have been largely
concerned with three related foundational questions which are often
intermingled, but which I believe should be kept apart:

1. A semi-empirical question: Is quantum mechanics complete? In other words,
do we have to supplement or restrict the formalism by some additional assumptions?

2. A mathematical-logical question: What are the constraints imposed by
quantum mechanics on its possible alternatives? This is where all the famous
"no-hidden-variables"\ theorems belong.

3. A philosophical question: Assuming that quantum mechanics is complete, what
then does it say about reality?

By \emph{quantum mechanics} I mean the Hilbert space formalism, including the
dynamical rule for the quantum state given by Schr\"{o}dinger's equation,
Born's rule for calculating probabilities, and the association of measurements
with Hermitian operators. These elements seem to me to be the core of the
(nonrelativistic) theory.

I shall be concerned mainly with the philosophical question. Consequently, for
the purpose of this paper the validity and completeness of the Hilbert space
formalism is assumed. By making this assumption I do not wish to prejudge the
answer to the first question. It seems to me dogmatic to accept the
completeness claim, since no one can predict what future theories will look
like. At the same time I think it is also dogmatic to reject completeness.
Present day alternatives to quantum mechanics, be they collapse theories like
GRW \cite{1}, or non-collapse theories like Bohm's \cite{2}, all suffer from
very serious shortcomings.

However, one cannot ignore the strong philosophical motivation behind the
search for alternatives. These are, in particular, two conceptual assumptions,
or perhaps dogmas that propel this search: The first is J. S. Bell's dictum
that the concept of measurement should not be taken as fundamental, but should
rather be defined in terms of more basic processes \cite{3}. The second
assumption is that the quantum state is a real physical entity, and that
denying its reality turns quantum theory into a mere instrument for
predictions. This last assumption runs very quickly into the measurement
problem. Hence, one is forced either to adopt an essentially non-relativistic
alternative to quantum mechanics (e.g. Bohm without collapse, GRW with it); or
to adopt the baroque many worlds interpretation which has no collapse and
assumes that all measurement outcomes are realized.

In addition, the first assumption delegates secondary importance to
measurements, with the result that the uncertainty relations are all but
forgotten. They are accepted as empirical facts, of course; but after
everything is said and done we still do not know why it is impossible to
measure position and momentum at the same time. In Bohm's theory, for example,
the commutation relations are adopted by fiat even on the level of individual
processes, but are denied any fundamental role in the theory.

My approach is traditional and goes back to Heisenberg, Bohr and von Neumann.
It takes the uncertainty relations as the centerpiece that demarcates between
the classical and quantum domain. This position is mathematically expressed by
taking the Hilbert space, or more precisely, the lattice of its closed
subspaces, as the structure that represents the "elements of reality" in
quantum theory. The quantum state is a derived entity, it is a device for the
bookkeeping of probabilities. The general outlook presented here is thus
related to the school of quantum information theory, and can be seen as an
attempt to tie it to the broader questions of interpretation. I strive to
explain in what way quantum information is different from classical
information, and , perhaps why.

The main point is that the Hilbert space formalism is a "logic of partial
belief" in the sense of Frank Ramsey \cite{4}. In such a logic one usually
distinguishes between possible "states of the world" (in Savage's terminology
\cite{5}), and the probability function. The former represent an objective
reality and the latter our uncertainty about it. In the quantum context
possible states of the world are represented by the closed subspaces of the
Hilbert space while the probability is derived from the $\left\vert
\psi\right\rangle $ function by Born's rule. In order to avoid confusion
between the objective sense of \emph{possible state} (subspace), and
$\left\vert \psi\right\rangle $- which is also traditionally called the state-
we shall refer to the subspaces as \emph{events}, or \emph{possible
events},\emph{\ }or \emph{possible outcomes }(of experiments). To repeat, my
purpose is to defend the position that the Hilbert space formalism is
essentially a new theory of probability, and to try to grasp the implications
of this structure for reality.

The initial plausibility of this approach stems from the observation that
quantum mechanics uses a method for calculating probabilities which is
different from that of classical probability theory\footnote{This position has
been expressed often by Feynman \cite{6}, \cite{7} . For more references, and
an analysis of this point see \cite{8}}. Moreover, in order to force quantum
probability to conform to the classical mold we have to add objects
(variables, events) and dynamical laws over and above those of quantum theory.
This state of affairs calls for a philosophical analysis because the theory of
probability is a theory of inference and, as such, is a guide to the formation
of rational expectations.

The relation between the above stated purpose and the completeness assumption
should be stressed again. We can always avoid the radical view of probability
by adopting a non-local, contextual hidden variables theory such as Bohm's.
But then I believe, the philosophical point is missed. It is like taking
Steven Weinberg's position on space-time in general relativity: There is no
non-flat Riemannian geometry, only a gravitational field defined on a flat
space-time that appears as if it gives rise to geometry \cite{9, 10, 11}. I
think that Weinberg's point and also Bohm's theory are justified only to the
extent that they yield new discoveries in physics (as Weinberg certainly
hoped). So far they haven't.

Jeffrey Bub was my thesis supervisor over a quarter of a century ago, and from
him I have Iearnt the mysteries of quantum mechanics and quantum logic
\cite{12}. For quite a while our attempts to grasp the meaning of the theory
diverged, but now seem to converge again \cite{13}. It is a great pleasure for
me to contribute to this volume in honor of a teacher and a dear friend.

\section{The event structure}

\subsection{Impossibility, certainty, identity, and the non contextuality of
probability}

Traditionally a theory of probability distinguishes between the set of
possible events \ (called the algebra of events, or the set of states of
Nature, or the set of possible outcomes) and the probability measure defined
on them. In the Bayesian approach what constitutes a possible event is
dictated by Nature, and the probability of the event represents the degree of
belief we attach to its occurrence. This distinction, however , is not sharp;
what is possible is also a matter of judgment in the sense that an event is
judged impossible if it gets probability zero in all circumstances. In the
present case we deal with physical events, and what is impossible is therefore
dictated by the best available physical theory. Hence, probability
considerations enter into the structure of the set of possible events. We
represent by $0$ the equivalence class of all events which our physical theory
declares to be utterly impossible (never occur, and therefore always get
probability zero) and by $1$ what is certain (always occur, and therefore get
probability one).

Similarly, the \emph{identity} of events which is encoded by the structure
also involves judgments of probability in the sense that \emph{identical
events always have the same probability. }This is the meaning of accepting a
structure as an algebra of events \emph{in a probability space}. An important
example is the following: Consider two measurements $A$, $B$, which can be
performed together, so that $[A,B]=0$; and suppose that $A$ has the possible
outcomes $a_{1},a_{2},...,a_{k}$, and $B$ the possible outcomes $b_{1}%
,b_{2},...,b_{r}$. Denote by $\{A=\nolinebreak a_{i}\}$ the event "the outcome
of the measurement of $A$ is $a_{i}$", and similarly for $\{B=\nolinebreak
b_{j}\}$. \ Now consider the identity:
\begin{equation}
\{B=b_{j}\}=%
{\displaystyle\bigcup\limits_{i=1}^{k}}
(\{B=b_{j}\}\cap\{A=a_{i}\}) \label{1}%
\end{equation}
This is the distributivity rule which holds in this case\ as it also holds in
all classical cases. This means, for instance, that if $A$ represents the roll
of a die with six possible outcomes and $B$ the flip of a coin with two
possible outcomes, then Eq (\ref{1}) is trivial. Consequently the probability
of the left hand side of Eq (\ref{1}) equals the probability of the right hand
side, for every probability measure.

In the quantum mechanical context this observation has further implications.
If $A$, $B$, $C$, are observables such that $[A,B]=0$, and $[B,C]=0$ but
$[A,C]\neq0$. Then the identity
\begin{equation}%
{\displaystyle\bigcup\limits_{i=1}^{k}}
(\{B=b_{j}\}\cap\{A=a_{i}\})=\{B=b_{j}\}=%
{\displaystyle\bigcup\limits_{i=1}^{l}}
(\{B=b_{j}\}\cap\{C=c_{i}\}) \label{2}%
\end{equation}
holds, where $c_{1},c_{2},...,c_{l}$ are the possible outcomes of $C$. By the
rule \emph{Identical events always have the same probability} we conclude that
the probabilities of all three expressions in Eq (\ref{2}) are equal. This is
\emph{the principle of the non-contextuality of probability}. There is a large
body of literature which attempts to justify this principle\footnote{The
terminology was introduced in \cite{14}. See also \cite{15}, and the criticism
by Stairs \cite{16}. In case no commitment is made regarding the lattice of
subspaces as an event structure, the non contextuality of probability requires
a special justification. For example, in the many worlds interpretation
\cite{17}, \cite{18}.
\par
{}}. For why should we apply the same probability to $\{B=b_{j}\}$ in the
$A,B$ context as in the $B,C$ context? If this is a good question in the
quantum domain it should be an equally good question in the classical regime.
For consider Eq \nolinebreak(\ref{1}) with $A$ representing the throw of a
die, and $B$ the flip of a coin. Now think of two contexts: In one we just
flip the coin without rolling the die; in the other we do both. Why should the
probability of $\{B=b_{j}\}$ be the same in both contexts? (regardless of our
judgment about the dependence, or independence of the events). By the very act
of putting the outcomes of the two procedures \textquotedblleft coin
flipping\textquotedblright\ and \textquotedblleft die
throwing\textquotedblright\ in the same \emph{probability} space (the product
space) we are ipso facto assuming Eq (\ref{1}) as an identity in a probability
space which implies equality of probabilities. Although routinely made, this
assumption ultimately represents an empirical judgment. Counterexamples are
hard to come by, and are usually quite contrived.

My proposal to take the Hilbert space formalism as a Ramsey type logic of
partial belief involves the same commitment. Hence, in the following I
\emph{assume }that the $0$ of the algebra of subspaces represents
impossibility (zero probability in all circumstances) $1$ represents certainty
(probability one in all circumstances), and the identities such as Eq
(\ref{1}) and Eq (\ref{2}) represent identity of probability in all
circumstances. This is the sense in which the lattice of closed subspaces of
the Hilbert space is taken as an algebra of events. I take these judgments to
be natural extensions of the classical case; a posteriori, they are all justified empirically.

\subsection{The axioms}

In their 1936 seminal paper \textquotedblleft The logic of quantum
mechanics\textquotedblright\ Garrett Birkhoff and John von Neumann \cite{19}
formulated the quantum logical program. Their strategy was to take the
following steps:

1. Identify the quantum structure which is the analogue of the event structure
of classical statistical mechanics.

2. Distill a set of principles underlying this structure and formulate them as axioms.

3. Show that the quantum structure is, in some sense, THE model of the axioms.

Birkhoff and von Neumann identified the quantum event structure (which they
called "quantum logic") as the algebra of closed subspaces of a Hilbert space.
In the rest of this section I shall review the efforts to accomplish steps 2
and 3 of their program, that is, begin with the axioms and generate the
structure. The elements in the structure we shall refer to as
\textquotedblleft events\textquotedblright, or \textquotedblleft
outcomes\textquotedblright\ (meaning outcomes of gambles or of measurements)
or\ sometimes loosely as "propositions" (meaning propositions that describe
the events). Notice that the axioms below are shared by both classical and
quantum systems, with the exception of the last axiom. It should also be noted
that I do not claim that this structure is \emph{logic} in the same sense that
the predicate calculus or intuitionistic logic are. (Nor do I think that
Birkhoff and von Neumann made such a claim).\footnote{The strong operational
approach of Finkelstein \cite{20}, and Putnam \cite{21} regarding the logical
connectives is -in the most charitable interpretation- a hidden variables
theory in disguise, see \cite{22}.} A proposition that describes a possible
event in a probability space is of a rather special kind. It is constrained by
the requirement that there should be a viable procedure to determine whether
the event occurs, so that a gamble that involves it can be unambiguously
decided. This means that we exclude many propositions. For example,
propositions that describe past events of which we have only a partial record,
or no record at all. We also exclude undecidable mathematical propositions
such as the continuum hypothesis, and many other propositions that form a part
of the standard conception of logic. Our structure is \textquotedblleft
logic\textquotedblright\ only insofar as it is the event component of a
\textquotedblleft logic of partial belief\textquotedblright.

We use small Latin letters $x,y,...$, to designate events, and denote by
$L$\ the totality of events. $\cap$\ stands for intersection,\ $\cup$\ for
union,\ and implication is denoted by $\leq$. Finally, $x^{\bot}$\ denotes the
complement of $x$. The certain event is denoted by $1$\ and the null event by
$0$.

These are the axioms:

\textbf{S1} $\ x\leq x$.

\textbf{S2} $\ $If $x\leq y$\ and $y\leq z$\ then $x\leq z$.

\textbf{S3} \ If $x\leq y$\ and $y\leq x$\ then $x=y$.

\textbf{S4} \ $0\leq x\leq1$

\textbf{S5} $\ x\cap y\leq x$, and $x\cap y\leq y$, and if $z\leq x$\ and
$z\leq y$\ then $z\leq$\ $x\cap y$.

\textbf{S6} \ $x\leq x\cup y$, and $y\leq x\cup y$, and if $x\leq z$\ and
$y\leq z$\ then $x\cup y\leq z$.

\textbf{O1} $(x^{\bot})^{\bot}=x$

\textbf{O2} $x\cap x^{\bot}=0$\ and $x\cup x^{\bot}=1$

\textbf{O3} $x\leq y$\ implies $y^{\bot}\leq x^{\bot}$.

\textbf{O4} $\ $Orthomodularity if $x\leq y$\ then $y=x\cup(y\cap x^{\bot}) $.

Axiom O4 is sometimes replaced by a stronger axiom:

\textbf{O4*} $\ $Modularity if $x\leq z$\ then $x\cup(y\cap z)=(x\cup y)\cap
z$.

The axioms \textbf{S1}-\textbf{S6}, \textbf{O1}-\textbf{O4} are true in the
classical system of propositional logic or, more precisely, in the
Lindenbaum-Tarski algebra of such a logic, when we interpret the operations as
logical connectives. The rest of the axioms are more specific to the physical context.

\textbf{H1} \ Atomism: If $x\lneqq y$\ then there is an atom $p$\ such that
$p\leq y$\ and $p\nleqslant x$. Here by an \emph{atom} we mean an element
$0\neq p\in L$\ such that $x\leq p$ entails $x=0$\ or $x=p$.

\textbf{H2} \ Covering property: For all atoms $p$\ and all elements $x$\ if
$x\cap p=0$\ then $x\leq y\leq x\cup p$\ entails $y=x$\ or $y=x\cup p$.

Atomism and the covering property are introduced to ensure that every element
of the lattice is a union of atoms. The atoms, whose existence is guaranteed
by \textbf{H1}, are maximally informative propositions. In the classical case
they correspond to points in the phase space (or rather, singleton subsets of
phase space); in the quantum case they correspond to one dimensional subspaces
of the Hilbert space.\footnote{At a later stage von Neumann gave up the
atomicity assumption. The reason has to do with the absence of a uniform
probability distribution over the closed subspaces of an infinite dimensional
Hilbert space. The non-atomic structures that resulted are his famous
continuous geometries, see \cite{23}.}

\textbf{H3} \ Completeness: if $S\subset L$\ then $\cup_{a\in S}a$\ and
$\cap_{a\in S}a$\ exist.

Usually we do not assume such a strong axiom in the classical physical case.
There, the algebra of possible events is the $\sigma$-algebra of Lebesgue
measurable subsets of phase space, which is assumed to be closed only under
countable unions and intersections. However, axiom \textbf{H3} is
\emph{consistent} with the classical physical event space. It is known that in
some models of set theory every set of reals is Lebesgue measurable \cite{24}.
In such models \textbf{H3} will automatically be satisfied for the Lebesgue
algebra in phase space. This means that no substantial difference between the
classical and quantum case arises from \textbf{H3}.

The one single axiom that separates the quantum from the classical domain is

\textbf{H4} \ Irreducibility: If $z$\ satisfies for all $x\in L$\ $x=(x\cap
z)\cup(x\cap z^{\bot})$\ then $z=0$\ or $z=1$.

This last axiom is non-classical in the following sense: there is only one
Boolean algebra which is irreducible, the trivial algebra $\{0,1\}$. In
classical physics the set of events is a large Boolean algebra. In fact, it is
\emph{totally reducible}:\ for all $x$\ and all $z$\ we have $x=(x\cap
z)\cup(x\cap z^{\bot})$.

So consider the case
\begin{equation}
x\neq(x\cap z)\cup(x\cap z^{\bot}) \label{3}%
\end{equation}

The intuitive meaning of Eq (\ref{3})\ is that the events $x$\ and $z$\ are
\emph{incompatible}, that is, cannot be the outcomes of a single experiment.
Thus, axiom \textbf{H4} is the formal expression of indeterminacy. Later we
shall see how Eq\nolinebreak(\ref{3}) entails a more familiar uncertainty
relation between the probabilities of $x$\ and $z$. For the sake of
illustration, at this stage, consider the case in which $x$ and $z$ are atoms.
One implication of Eq(\ref{3}) is that \emph{there are non orthogonal atoms}.
So consider some measurement in which $x$ is the \emph{actual} outcome, and
the other possible outcomes are $x^{\prime}$, $x^{\prime\prime}$,...etc., all
orthogonal to $x$, so that $z$ is not among them. This means that after the
measurement is performed we gain no knowledge as to whether $z$ is the case or
not. This state of affairs would not be very surprising were it not for the
fact that $x$ and $z$ are atomic events; but in this case it seems to imply
that \emph{there is no fact of the matter as to whether }$z$\emph{\ is the
case or not}. In other words, no certain record about the possible outcome $z$
is obtainable, in principle, while we perform the $x$ measurement. By "fact" I
mean here, and throughout, a \emph{recorded} fact, an actual outcome of a
measurement. Restricting the notion of \textquotedblleft
fact\textquotedblright\ in this way should not be understood, at this stage,
as a metaphysical thesis about reality. It is simply the concept of
\textquotedblleft fact\textquotedblright\ that is analytically related to our
notion of \textquotedblleft event\textquotedblright, in the sense that only a
recordable event can potentially be the object of a gamble. Later, in section
4.1 and in the last section we shall come back to this issue, when we discuss
the implications of the theory to the structure of reality.

\subsection{Representations and the gap}

In the classical case we assume that \emph{for all }$x$\emph{\ and }%
$z$\emph{\ the following holds} $x=(x\cap z)\cup(x\cap z^{\bot})$. This makes
the lattice $L$ an atomic Boolean algebra. More specifically $(L,0,1,\leq
,\cap,\cup,\bot)$\ is isomorphic to the Boolean algebra of the subsets of the
set of all atoms, with the the usual Boolean operators, with $1$\ the set of
all atoms and $0$\ the null set.

The representation theorem for quantum systems is more complicated, in this
case\ $(L,0,1,\leq,\cap,\cup,\bot)$\ is isomorphic to the lattice of subspaces
of a vector space with a scalar product, more specifically:

1. There is a division ring $K$\ (field whose product is not necessarily
commutative), with involutional automorphism $\ast:K\rightarrow K$,\ that is,
for all $\alpha,\beta\in K$\ $\ \alpha^{\ast\ast}=\alpha$, $(\alpha
+\beta)^{\ast}=\alpha^{\ast}+\beta^{\ast}$\ , $(\alpha\beta)^{\ast}%
=\beta^{\ast}\alpha^{\ast\text{ }}$

2. There's a (left) vector space $V$\ over $K$.

3. There's a Hermitian form $<,>:V\times V\rightarrow K$\ satisfying for all
$u,v,w\in V$, and $\alpha,\beta\in K$

$<\alpha u+\beta v,w>=\alpha<u,v>+\beta<v,w>$,

$<u,\alpha v+\beta w>=<u,v>\alpha^{\ast}+<u,w>\beta^{\ast}$,

$<u,v>=<v,u>^{\ast}$,

$<u,u>=0$\ if and only if $u=0$.

Let $X\subset V$\ be a subspace, let $X^{\bot}=\{v\in V;<u,v>=0\ \forall u\in
X\}$\ then $X^{\bot}$\ is also a subspace. If $X=X^{\bot\bot}$\ \ we shall say
that $X$\ is \emph{closed}, then we have $V=X\oplus X^{\bot}$. The
representation theorem asserts that $L$\ is isomorphic to the lattice of
closed subspaces of $V$, in other words $L\simeq\{X\subset V;X=\nolinebreak
X^{\bot\bot}\}$. The operation $\cap$\ is just subspace intersection, and
$X\cup Y=(X^{\bot}\cap Y^{\bot})^{\bot}$.

The proof of this representation theorem has essentially two parts. The first
is the classical representation theorem for projective geometries which goes
back to the middle of the 19th century.\footnote{By M\"{o}bius and von Staudt.
For the standard geometric construction see \cite{25}. A modern account which
stresses the algebraic aspects is in Artin's classic \cite{26}} An
irreducible, atomic, complete, lattice with a complementation $\bot$ that
satisfies \textbf{O2}, and which is modular (\textbf{O4*}) is a projective
geometry. The traditional result on the coordinatization of projective
geometries yields the field $K$ and the vector space $V$ over it. Adding the
stronger conditions on $\bot$, in particular \textbf{O4}, enabled Birkhoff and
von Neumann to derive the inner product structure on $V$. Note that so far we
have not introduced any explicit physical assumption, or even a probabilistic
assumption, save perhaps the indeterminacy implicit in \textbf{H4}.
Nevertheless, we see that the principle of superposition (that is, the fact
that $V$ is a linear space) already presents itself.

In both the classical and the quantum cases some additional assumptions are
needed to obtain the actual models. In the quantum case the construction will
be completed if we are able to infer that $K=\mathbb{C}$\ (the field of
complex numbers) on the basis of a probabilistic or a physically intuitive
axiom. At least we would like to force $K$ to be either the field of real
numbers, or the complex numbers, or the quaternions. In these cases the inner
product of a non-zero vector by itself $<u,u>$ is a positive real number, and
Gleason's theorem describes the probabilistic structure. This is a gap in the
argument which has been closed to a certain extent (in the case of infinite
dimensional Hilbert spaces) by the work of Sol\`{e}r \cite{27, 28}. It is
hoped that a reasonable more straightforward probabilistic or information
theoretic axiom (such as a constraint on tensor-like products) will close the
gap even more tightly .

\subsection{Sol\`{e}r's axiom and theorem}

The best result known in this direction involves a geometric axiom. It is the
celebrated theorem of Maria Pia Sol\`{e}r which applies in case the lattice is
infinite dimensional. The extra axiom connects a projective geometric concept
(harmonic conjugation) to the orthogonality structure. Recall that a
projective geometry is associated with the lattice in the following way: Every
atom is a \emph{point} every pair of atoms generates \emph{a projective line}
and every triple of atoms which are not colinear determine a \emph{projective
plane}. Let $x$\ and $y$ be two atoms then the line through them is $x\cup y$.
Suppose that $z$ is another atom on this line, $z\leq x\cup y$, then we
construct a fourth point $w\leq x\cup y$ on the line which is called the
harmonic conjugate of $z$ relative to $x$ and $y$ -denoted by $w=\mathcal{H}%
(z;x,y)$- as follows (Fig \ref{1}): Let $u\nleq x\cup y$ be arbitrary and let
$v\leq x\cup u$, $v\neq x,u$. Denote by $s=\left(  z\cup u\right)  \cap\left(
y\cup v\right)  $ and $t=(x\cup s)\cap\left(  z\cup v\right)  $ then
$w\doteqdot\mathcal{H}(z;x,y)\doteqdot(u\cup t)\cap(x\cup y)$. The harmonic
conjugate is unique (that is, independent of the choice of $u$ and $v$). It is
a basic construction in projective geometry, closely related to the
definitions of the algebraic operations in the field $K$ (realized as the
projective line).%
\begin{figure}
[ptb]
\begin{center}
\includegraphics[
natheight=1.599900in,
natwidth=2.559800in,
height=1.6362in,
width=2.6013in
]%
{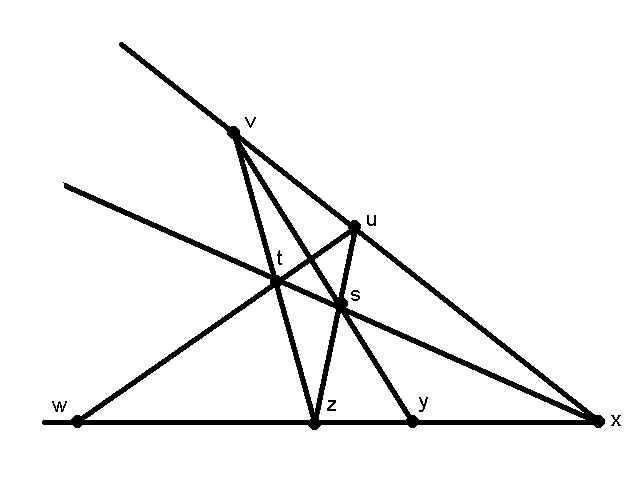}%
\caption{Harmonic conjugation}%
\end{center}
\end{figure}

Soler's axiom may be phrased as follows:

\textbf{SO \ }If $x$ and $y$ are orthogonal atoms then there is $z\leq x\cup
y$ such that $w=\mathcal{H}(z;x,y)$ is orthogonal to $z$. In other words,
$\mathcal{H}(z;x,y)=z^{\bot}\cap(x\cup y)$.

Intuitively, such a $z$ bisects the angle between $x$ and $y$, that is,
defines $\sqrt{2}$ in the field $K$. Soler's proved

\begin{theorem}
\emph{If }$L$\emph{\ }is infinite dimensional and satisfies\emph{\ \textbf{SO}
}then\emph{\ }$K$\emph{\ }is\emph{\ }$\mathbb{R}$\emph{\ }or\emph{\ }%
$\mathbb{C}$\emph{\ }or the quaternions.
\end{theorem}

In fact she proved a stronger result, assuming only that there is an infinite
sequence of orthogonal atoms $\{x_{i}\}_{i\in\mathbb{N}}$ such that $x=x_{i}$
and $y=x_{i+1}$ satisfy \textbf{SO} for every $i=1,2,..$. The axiom
\textbf{SO} may be given a probabilistic interpretation in the spirit Ramsey
as we shall see subsequently.

\section{Probability Measures: Gleason's theorem, Wondergraph and Sol\`{e}r's
axiom.}

\subsection{Gleason's theorem}

Assume that the set of possible events (or possible measurement outcomes, or
propositions) is the lattice $L=L(\mathbb{H)}$ of subspaces of a \emph{real or
complex} Hilbert space $\mathbb{H}$. For simplicity, we shall concentrate on
the finite dimensional case. Our aim is to tie this structure to
probabilities, and by doing so to provide further evidence that the elements
of $L$ can be seen as representing quantum events. Moreover, we shall see how
the traditional features and "paradoxes" of quantum mechanics are expressed
and resolved in the quantum probabilistic language.

First a few words to connect measurements and outcomes in the more traditional
view with the present notations. Here we shall be concerned with measurements
that have a finite set of possible outcomes. Let $A$ be an observable (a
Hermitian operator) with $n$ distinct possible numerical real values (the
eigenvalues of $A$) $\alpha_{1,}\alpha_{2,...,}\alpha_{n}$. With each value
corresponds an event $x_{i}=\{A=\alpha_{i}\}$ meaning \textquotedblleft the
outcome of a measurement of $A\ $is$\ \alpha_{i}$\textquotedblright\ We
identify this event with the subspace of $\mathbb{H}$ spanned by the
eigenvectors of $A$ having the eigenvalue $\alpha_{i}$. The events $x_{i}$ are
pair-wise orthogonal elements of $L$. The sub lattice that $x_{1}%
,x_{2},...,x_{n}$ generate is a finite Boolean algebra which we shall denote
by $\mathcal{B}=\left\langle x_{1},x_{2},...,x_{n}\right\rangle $. In case $n$
is the dimension of the space $\mathbb{H}$ each one of the events $x_{i}$ is
an atom and the observable $A$ is said to be maximal.

\emph{Subsequently we shall identify any observable }$A$\emph{\ with the
Boolean algebra }$\left\langle x_{1},x_{2},...,x_{n}\right\rangle
$\emph{\ generated by its outcomes. }Note that this is an unusual
identification. It means that we equate the observables $A$ and $f(A)$,
whenever $f$ is a one-one function defined on the eigenvalues of $A$. This
step is justified since we are interested in \emph{outcomes} and not their
labels, and hence in such a \textquotedblleft scale free\textquotedblright%
\ concept of observable. (It is like replacing the numbers $1,2,...,6$ on the
face of a die by the numbers $2,3,...,7$ respectively). The converse is also
true, with each orthogonal set of elements $x_{1},x_{2},...,x_{n}$ of $L$
there corresponds an observable whose eigenspaces include these elements.

Probability measures which are definable on $L$ were characterized many years
ago in case $n=\dim\mathbb{H\geq}3$. Since every set of $n$ orthogonal atoms
represents the outcomes of a possible measurement, and since they are all the
possible outcomes we are motivated to introduce

\begin{definition}
Suppose that $\mathbb{H}$ is of a finite dimension $n$ over the complex or
real field. A real function $P$ defined on the atoms in $L$ is called a state
(or alternatively, a probability function) on\emph{\ }$\mathbb{H}$ if the
following conditions hold
\end{definition}

\emph{1. }$P(0)=0$\emph{, and} $P(y)\geq0$\emph{\ for every element }$y\in L
$\emph{.}

\emph{2. If }$x_{1},x_{2},...,x_{n}$\emph{\ is an orthogonal set of atoms then
}$\sum_{j=1}^{n}P(x_{j})=1$\emph{.}

The probability of every lattice element $y\in L$ is then fixed since it is a
union of a set of orthogonal atoms $y=x_{1}\cup...\cup x_{r}$, so that
$P(y)=\sum_{j=1}^{r}P(x_{j})$. A complete description of the possible states
is given by Gleason's theorem \cite{29}:

\begin{theorem}
Given a state $P$ on a space of dimension $\geq3$ there is an Hermitian, non
negative operator $W$\ on $H$, whose trace is unity, such that $P(x)=<\overset
{\rightarrow}{x},W\overset{\rightarrow}{x}>$\ for all atoms $x\in L $, where
$<,>$\ is the inner product, and $\overset{\rightarrow}{x}$ is a unit vector
along $x$. In particular, if some $x_{0}\in L$ satisfies $P(x_{0})=1$ then
$P(x)=\left\vert <\overset{\rightarrow}{x}_{0},\overset{\rightarrow}%
{x}>\right\vert ^{2}$ for all $x\in L$ (Born's rule).
\end{theorem}

With the obvious conditions on convergence the above definition and theorem
generalize to the infinite dimensional case. The remarkable feature exposed by
Gleason's theorem is that the event structure dictates the quantum mechanical
probability rule. It is one of the strongest pieces of evidence in support of
the claim that the Hilbert space formalism is just a new kind of probability
theory. The quantum structure is in this sense much more constrained than the
classical formalism. The structure of the phase space of a classical system
does not gratly restrict the type of probability measures that can be defined
on it. The probability measures which are actually used in classical
statistical mechanics are introduced mostly by fiat or, in any case, are very
hard to justify.

G\"{o}del \cite{30} said in a different context : \textrm{"A probable decision
about the truth [of a new axiom] is possible ... inductively by studying its
\textquotedblleft success\textquotedblright. Success here means fruitfulness
in consequences in particular \textquotedblleft verifiable\textquotedblright%
\ consequences, i.e., consequences demonstrable without the axiom". }Importing
this insight from the mathematical domain to the present physical domain we
can see how the set of axioms for the structure, most of which are shared with
classical probability, give rise to the quantum mechanical probabilistic
structure which is otherwise left a mystery.

\subsection{Finite gambles and uncertainty}

So far we have dealt with the lattice $L$ in its entirety, and with everywhere
defined probability functions. The standard conceptions of Bayesian
probability theory make do, initially at least, of finite probability spaces.
A canonical situation handled by this theory is that of a gamble. In the words
of Ramsey: \textquotedblleft The old-established way of measuring a person's
belief \textquotedblright\ by proposing a bet, and seeing what are the lowest
odds which he will accept, is \textquotedblleft fundamentally
sound\textquotedblright\ \cite{4}. Our gambles will likewise be finite and
consist of four steps

1. A \textit{single} physical system is prepared by a method known to everybody.

2. A \textit{finite} set $\mathcal{M}$ of incompatible measurements, each with
a finite number of possible outcomes, is announced by the bookie. The agent is
asked to place bets on the possible outcomes of each one of them.

3. One of the measurements in the set $\mathcal{M}$ is chosen by the bookie
and the money placed on all other measurements is promptly returned to the agent.

4. The chosen measurement is performed and the agent gains or looses in
accordance with his bet on that measurement.

There are two reasons to concentrate on finite gambles of this kind. First, to
avoid over idealization; for it is hard to imagine someone betting on the
outcomes of all possible measurements (perhaps writing an IOU for each one of
them). Secondly, and more importantly, the infinite idealization blurs the
important fact that indeterminacy, and all other "strange" results associated
with quantum theory, are fundamentally combinatorial. The non-classical
behavior of the probabilities is already forced by a finite number of events
and the relations among them.

Recall that each measurement is identified with the Boolean algebra generated
by its possible outcomes in $L$: $\mathcal{B}=\left\langle x_{1}%
,x_{2},...,x_{m}\right\rangle $ \ (the\ $x_{i}$'s may not be atomic in
case\ $\mathcal{B}$ is not a maximal measurement). So a gamble $\mathcal{M}$
is just a set of such algebras $\mathcal{M=\{B}_{1},\mathcal{B}_{2}%
,...,\mathcal{B}_{k}\}$. We do not assume that the gambler knows quantum
theory. All she is aware of is the logical structure which consists of these
sets of outcomes. In particular, \emph{she recognizes identities}, and the
cases where the same outcome is shared by more than one experiment. By acting
according to the standards of rationality the gambler will assign
probabilities to the outcomes. To see this, assume that $P(x\mid\mathcal{B}) $
is the probability assigned by the agent to the outcome $x$ in measurement
$\mathcal{B}$, where $\mathcal{B}\in\mathcal{M}$ and$\ x\in$ $\mathcal{B}$.

\textbf{RULE\ 1:} \textit{For each measurement }$\mathcal{B}\in\mathcal{M}$
\textit{the function }$\ P(\cdot\mid\mathcal{B})$\textit{\ is a probability
distribution on }$\mathcal{B}$.

This follows directly from classical probability theory. Recall that after the
third stage in the quantum gamble the agent faces a bet on the outcome of a
single measurement. The situation at this stage is essentially the same as a
tossing of a coin or a casting of a die. Hence, the probability values
assigned to the possible outcomes of the chosen measurement should be coherent.

\textbf{RULE 2:} \textit{If }$\mathcal{B}_{1},\mathcal{B}_{2}\in\mathcal{M}%
$,$\ $and $y\in\mathcal{B}_{1}\cap\mathcal{B}_{2}$\textit{\ then }%
$P(y\mid\mathcal{B}_{1})=P(y\mid\mathcal{B}_{2})$.

The rule asserts the non contextuality of probability, discussed in section
2.1. Suppose that $\mathcal{B}_{1}=\left\langle x_{1},x_{2},...,x_{m}%
\right\rangle $ and $\mathcal{B}_{2}=\left\langle z_{1},z_{2},...,z_{r}%
\right\rangle $ then $y\in\mathcal{B}_{1}\cap\mathcal{B}_{2}$ implies that
$(x_{1}\cap y)\cup...\cup(x_{m}\cap y)=y=(z_{1}\cap y)\cup...\cup(z_{r}\cap
y)$. Rule 2, therefore, follows from this identity between events, and the
principle that identical events in a probability space have equal probabilities.

To take the discussion closer to the lattice theoretic conception consider
finite subsets of events, $\Gamma\subset L$.

\begin{definition}
Two propositions $x$ and $y$ of $\Gamma$ are \emph{compatible} if $x=(x\cap
y)\cup\nolinebreak(x\cap y^{\bot})$ and $y=(y\cap x)\cup(y\cap x^{\bot}) $. A
state (or probability function) $\Gamma$ is a real function $P$ on $\Gamma$
such that
\end{definition}

\emph{a. }$P(x)\geq0$\emph{\ for all }$x\in\Gamma$

\emph{b. }$P(x^{\bot})=1-P(x)$\emph{\ whenever }$x,x^{\bot}\in\Gamma$

\emph{c. }$P(x\cup y)+P(x\cap y)=P(x)+P(y)$\emph{\ whenever }$x$\emph{\ and
}$y$\emph{\ are compatible and }$x,y\in\Gamma$\emph{.}

Such probability functions defined over finite subsets of events in the
lattice are the subject of our study. Note that we do not put any requirements
on such $P$'s apart from the three conditions a, b, c, in the definition. In
particular, probability functions on $\Gamma$ are not constrained to be
induced by quantum mechanical states. The relation between this definition and
the gambles introduced previously is clear. Given any gamble $\mathcal{M}$ as
above the set of events is $\Gamma=\mathcal{B}_{1}\cup\mathcal{B}_{2}%
\cup...\cup\mathcal{B}_{k}\subset L$. Every probability function which follows
\textbf{RULE 1} and \textbf{RULE 2 }satisfy the conditions a, b, c, in
definition 2.

As a simple example which demonstrates an uncertainty relation consider the
following quantum gamble $\mathcal{M}$ consisting of seven incompatible
measurements (Boolean algebras), each generated by its three possible atomic outcomes:%

\begin{align*}
&  \left\langle x_{1},x_{2},y_{2}\right\rangle ,\ \left\langle x_{1}%
,x_{3},y_{3}\right\rangle ,\ \left\langle x_{2},x_{4},x_{6}\right\rangle
,\ \left\langle x_{3},x_{5},x_{7}\right\rangle ,\smallskip\\
&  \ \left\langle x_{6},x_{7},y\right\rangle ,\ \left\langle x_{4},x_{8}%
,y_{4}\right\rangle ,\ \left\langle x_{5},x_{8},y_{5}\right\rangle
\end{align*}
Note that some of the outcomes are shared by two measurements, these are
denoted by the letter $x$. The other outcomes each belong to a single algebra,
and are denoted by a $y$. The orthogonality relations among the generators are
depicted in the \textit{orthogonality graph} in Fig(\ref{2}), which is a part
of Kochen and Specker's famous "cat's cradle"\cite{31}. Each node in the graph
represents an outcome, two nodes are connected by an edge if, and only if the
corresponding outcomes belong to a common Boolean algebra (measurement); each
triangle represents the generators of one of the Boolean algebras.%
\begin{figure}
[ptb]
\begin{center}
\includegraphics[
natheight=3.072700in,
natwidth=4.780700in,
height=3.1176in,
width=4.8334in
]%
{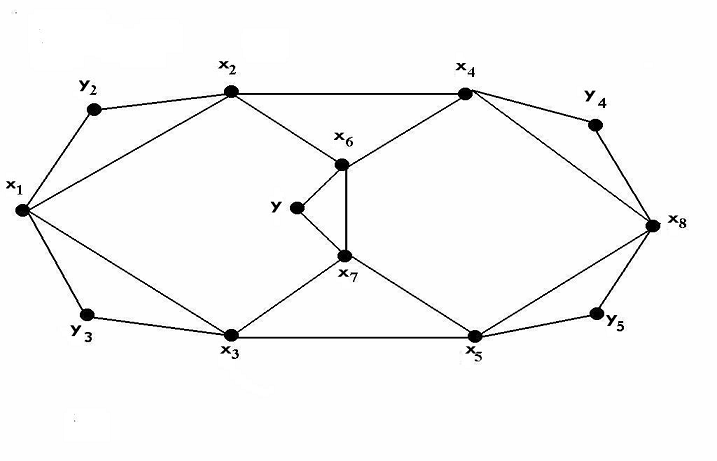}%
\caption{Cat's cradle}%
\end{center}
\end{figure}

The probabilities of each triple of outcomes of each measurement should sum to
1, for example, $P(x_{4})+P(x_{8})+P(y_{4})=1$. There are altogether seven
equations of this kind. Combining them with the fact that probability is
non-negative it is easy to prove that the probabilities assigned by our
rational agent should satisfy $P(x_{1})+P(x_{8})\leq\frac{3}{2}$ \cite{15}.
This is an example of an \textit{uncertainty relation}, a constraint on the
probabilities assigned to the outcomes of incompatible measurements. In
particular, if the system is prepared in such a way that it is rational to
assign $P(x_{1})=1$ then the rules of quantum gambles force $P(x_{8})\leq
\frac{1}{2}$.

This result is a special case of a more general principle given by \cite{32,
33}

\begin{theorem}
(logical indeterminacy principle) Assuming $\dim\mathbb{H}\geq3,$ let $x$ and
$y$ \textit{be two incompatible atoms in the lattice }$L=L(\mathbb{H)}$, that
is, $x\neq(x\cap y)\cup(x\cap y^{\bot})$. Then there is a finite set
$\Gamma\subset L\mathbb{(H})$ with $x,y\in\Gamma$ such that every state $P $
on $\Gamma$ satisfies $P(x)+P(y)<2$. In fact we have more: $P(x),P(y)$
$\in\{0,1\}$ if and only if $P(x)=P(y)=0$.
\end{theorem}

This theorem explains the sense in which axiom \textbf{H4 }-the axiom of
irreducibility- expresses indeterminacy. This axiom asserts that \emph{for
every non trivial }$x$\emph{\ there is a }$y$\emph{\ such that }$x\neq(x\cap
y)\cup(x\cap y^{\bot})$. By the logical indeterminacy principle the
probability value of at least one of the events $x$ or $y$ must be strictly
between zero and one, unless they both have probability zero. Moreover, this
fact is already forced by the relation between $x$, $y$ and finitely many
other events. Remember also that \textbf{H4} is the only axiom (except
\textbf{SO}) that distinguishes between the classical and quantum structures.

\subsection{Wondergraph}

The previous theorem is typical in the sense that all features of quantum
probability, even the quantitative features, can be forced by the logical
relations among finitely many events. This follows from a construction of a
particular finite set of atoms in $\mathbb{R}^{3}$ which, together with the
orthogonality relations among its elements will be called the
\emph{Wondergraph}.

Let us introduce first the notion of \emph{a} \emph{frame function} which
generalizes the concept of a state

\begin{definition}
Let $\Gamma\subseteq L(\mathbb{H})$ be a set of atoms of $L(\mathbb{H})$ where
$\dim\mathbb{H}=n$. A frame function on $\Gamma$ is a real function $f$ on
$\Gamma$ such that all orthogonal sets of atoms $x_{1},x_{2},...,x_{n} $ in
$\Gamma$ satisfy $\sum_{j=1}^{n}f(x_{j})=C$; where $C$ is a constant.
\end{definition}

Consider the case of $\mathbb{R}^{3}$ the smallest space to which Gleason's
theorem applies. Let $\overrightarrow{e_{1}}=(1,0,0)$, $\overrightarrow{e_{2}%
}=(0,1,0)$ and $\overrightarrow{e_{3}}=(0,0,1)$ be the standard basis in
$\mathbb{R}^{3}$ and $\overrightarrow{b_{ij}}=\frac{1}{\sqrt{2}}%
(\overrightarrow{e_{i}}+\overrightarrow{e_{j}})$, $1\leq i<j\leq3$. Denote by
$e_{i}$ and $b_{ij}$ the one dimensional subspaces along these vectors. The
following theorem turns out to be equivalent to Gleason's theorem \cite{33}:

\begin{theorem}
(Wondergraph theorem) For every atom $z\in L(\mathbb{R}^{3})$ there is a
finite set of atoms $\Omega(z)\subset L(\mathbb{R}^{3})$ such that
$e_{i},b_{ij},z\in\Omega(z)$ and such that every frame function $f$ on
$\Omega(z)$ which satisfies $f(e_{i})=f(b_{ij})=0$, and $\left\vert
f(x)\right\vert \leq1$ for all $x\in\Omega(z)$ necessarily also satisfies
$\left\vert f(z)\right\vert \leq\frac{1}{2}$. Moreover, $\left\vert
\Omega(z)\right\vert $, the number of elements of $\Omega(z)$, is \ the same
for all $z$.
\end{theorem}

Note that the condition $f(e_{i})=0,$ $\ 1\leq i\leq3$ for the frame function
$f$ on $\Omega(z)$ entails that $f(x)+f(x^{\prime})+f(x^{\prime\prime})=0$ for
all orthogonal triples $x,x^{\prime},x^{\prime\prime}\in\Omega(z)$. To see why
Wondergraph theorem entails Gleason's theorem consider first

\begin{lemma}
Gleason's theorem for $\mathbb{R}^{3}$is true if and only if every bounded
frame function $f$ defined on the atoms of $L(\mathbb{R}^{3})$ which satisfies
$f(e_{i})=f(b_{ij})=0$ is identically zero.
\end{lemma}

The proof of the lemma is straightforward. It follows from the fact that the
quadric form $\left\langle \overrightarrow{x},A\overrightarrow{x}\right\rangle
$ induced by a self adjoint operator $A$ on $\mathbb{R}^{3}$ is uniquely
determined by the six \ numbers $\left\langle \overrightarrow{e_{i}%
},A\overrightarrow{e_{i}}\right\rangle $, $\left\langle \overrightarrow
{b_{ij}},A\overrightarrow{b_{ij}}\right\rangle $. Now, to see how Gleason's
theorem follows from Wondergraph let $f$ be a \emph{bounded} frame function
defined on the atoms of $L(\mathbb{R}^{3})$ which satisfies $f(e_{i}%
)=f(b_{ij})=0$. Normalize $f$ so that $\left\vert f(x)\right\vert \leq1$ for
all $x$. Take $z$ to be arbitrary, then the restriction of $f$ to $\Omega(z)$
is a frame function on $\Omega(z)$ and therefore $\left\vert f(z)\right\vert
\leq\frac{1}{2}$. Suppose the atoms of $\Omega(z)$ are $x_{1},...,x_{s}$ and
consider the set $\Omega_{1}(z)=\bigcup_{j=1}^{s}\Omega(x_{j})$. The
restriction of $f$ to $\Omega_{1}(z)$ is a frame function on each one of the
$\Omega(x_{j})$'s. Hence, $\left\vert f(x_{j})\right\vert \leq\frac{1}{2}$ for
all $x_{j}\in\Omega(z) $ and therefore $\left\vert f(z)\right\vert \leq
\frac{1}{4}$. Iterating this process we get that $\left\vert f(z)\right\vert $
becomes as small as we wish. Since $z$ is arbitrary the theorem follows.
Gleason's theorem for any Hilbert space follows from the case of
$\mathbb{R}^{3}$, as Gleason himself showed. Another way to extend the theorem
from $\mathbb{R}^{3}$ to higher real or complex dimensions is to construct
Wondergraphs in every (finite) dimension; which can be done once the three
dimensional real case is given.

The proof that Gleason's theorem entails the existence of Wondergraph is based
on model theory. As a part of the proof one also concludes that there is a
known algorithm to construct Wondergraph. The setback is that this algorithm
runs very slowly (it is, in fact , the decision algorithm for the theory of
real closed fields, which in the worst case runs in doubly exponential time).
Thus we pose a

\begin{problem}
Construct Wondergraph explicitly.
\end{problem}

Wondergraph allows one to reduce all the interesting quantum phenomena to
relations among finitely many events. This follows from:

\begin{corollary}
Given a finite set of atomic events $\Gamma_{0}$\emph{\ }and a real
number\emph{\ }$\varepsilon>0$\emph{\ }there is a finite set of atoms\emph{\ }%
$\Gamma$\emph{\ }such that
\end{corollary}

\emph{a. }$\ \Gamma_{0}\subset\Gamma$\emph{, the number of elements
}$\left\vert \Gamma\right\vert $\emph{\ of }$\Gamma$\emph{\ depends on
}$\varepsilon$\emph{\ and on }$\left\vert \Gamma_{0}\right\vert $\emph{\ but
not on the elements of }$\Gamma_{0}$\emph{.}

\emph{b. \ If }$P$\emph{\ is a state on }$\Gamma$\emph{\ then there is a
quantum state }$W$\emph{\ (non negative Hermitian operator with trace }%
$1$\emph{) such that}%

\[
\left\vert P(x)-<\overset{\rightarrow}{x},W\overset{\rightarrow}%
{x}>\right\vert <\varepsilon\quad for\ all\ x\in\Gamma_{0}%
\]

\emph{c. There is an algorithm to generate }$\Gamma$\emph{\ given }$\Gamma
_{0}$\emph{\ and }$\varepsilon$\emph{.}

For many of the famous "paradoxes" of quantum mechanics explicit constructions
of the required finite set $\Gamma$ exist \cite{15, 32, 33}. These include the
EPR-Bell argument, the Kochen and Specker theorem, and also generalizations of
Kochen and Specker to any given finite number of colors.

On a more fundamental level the importance of these results lies in the way
probabilities are associated with $L$, the algebra of all the possible
outcomes of all possible measurements. Remember that in the epistemic
conception of probability a \textquotedblleft fundamentally
sound\textquotedblright\ method of measuring a person's belief is
\textquotedblleft by proposing a bet and seeing what are the lowest odds he
will accept\textquotedblright. In order to fit the infinite structure $L$ into
this view of probability (or any other of the standard Bayesian accounts) we
consider only finite segments of $L$ and the probability functions definable
on them. These are the quantum gambles considered above. They are the
equivalents of classical gambles with dice, roulettes and cards. Some real
experiments involve arrangements which are like our gambles: A laboratory
device is prepared in such a way that it can perform either one of a few
incompatible measurements. Then, the experiment which is actually performed is
chosen at random. This gives quantum probability an "operational" flavour and,
hopefully removes some of the mystery connected with it, typically expressed
by words like "interference" and "superposition".

Another way to see this point is to think about the classical propositional
calculus. The Lindenbaum-Tarski algebra on countably many generators gives us
all the expressive power we need as far as the propositional connectives are
concerned. However, in practice we interpret (assign truth values) only to
finite subsets. By analogy, if we take $L$ as representing a "syntax"
encompassing symbols for all possible outcomes of all possible measurements,
then the "semantics" is the assignment of probability values to finite
sections of $L$. Gleason's theorem, in its Wondergraph version, implies that
this "semantics" is, in fact, complete:

\begin{corollary}
(Completeness) Suppose that an agent assigns probability values $P(x)$ to the
elements $x$ of a finite $\Gamma_{0}\subset L$, in a way that contradicts all
possible quantum assignments. Then there is a finite $\Gamma\supset\Gamma_{0}%
$, such that $P$ cannot be extended from $\Gamma_{0}$ to $\Gamma$. Hence, in a
larger gamble the agent can be shown to be irrational.
\end{corollary}

\subsection{Sol\`{e}r's axiom revisited}

Let us return to our axiomatic system and the axiom that closes the gap.
Recall that Sol\`{e}r's axiom asserts that for every pair of orthogonal atoms
$x$ and $y$ there is another atom $z$ in the plane they span, which bisects
the angle between $x$ and $y$. More formally: $\mathcal{H}(z;x,y)$, the
harmonic conjugate of $z$ with respect to $x$ and $y$, is orthogonal to $z$.

Assume that $L$ is infinite dimensional. In this case Sol\`{e}r's theorem,
when coupled with Gleason's theorem, implies that for any (globally defined)
state $P$ on $L$ and atoms $x,y\in L$, if $P(x)=1$ and $P(y)=0$ then
necessarily $x\bot y$, and there is an atom $z\leq x\cup y$ such that
$P(z)=\frac{1}{2}$. In other words, there is a precise interpolation between
probabilities zero and one.

The axiomatic systems of Bayesian probability theory typically include axioms
which imply interpolation of probability values. The most famous (or infamous)
one is Ramsey's axiom on the existence of an "ethically neutral" proposition
whose probability is one half (axiom 1 in Ramsey's system \cite{4}). The axiom
allows Ramsey to construct his theory of utilities (or "values", in his
terminology). Savage \cite{5}, who wanted to avoid notions like "ethical
neutrality", nevertheless also needs an interpolation principle for
probabilities, and assumes the existence of arbitrarily refined partitions.
This implies that one can obtain propositions with probabilities arbitrarily
close to any rational in the interval $[0,1]$.

I propose to read Sol\`{e}r's axiom as a probability interpolation axiom; or
at any rate to reformulate or replace it by a direct axiom about
probabilities. This, however, cannot be straightforward. We are not even
guaranteed that a globally defined state exists on $L$ in the first place.
However, we can use the fact that certain \emph{finite} orthogonality graphs
such as $\Gamma$ of theorem \nolinebreak3 force any state defined on them to
interpolate probability values between zero and one. This is our \emph{logical
indeterminacy principle }which expresses probabilistically the basic principle
that differentiates the quantum event structure from the classical one. Now,
we can turn the tables and assert axiomatically that orthogonality relations
like those in $\Gamma$ are realizable in $L$. This assertion indirectly
expresses the indeterminacy relations in their probabilistic sense. Here, for
example, is how this can be done:

Consider $L(\mathbb{R}^{3})$ and the rays $x$, $z$ through the
vectors:$\overrightarrow{x}=(1,0,0)$, and $\overrightarrow{z}=(1,1,0)$
respectively. Let $\Gamma=\Gamma(x,z)\subset L(\mathbb{R}^{3})$ be the finite
subset of rays guaranteed in theorem 3 (and explicitly constructed in
\cite{31,32}). This means that if $P$ is a state on $\Gamma$ with $P(x)=1$
then $0<P(z)<1$. Now, consider the rays in $\Gamma$ and their orthogonality
relations \emph{abstractly}, that is, as a graph, which we shall also denote
$\Gamma$. A candidate to replace Sol\`{e}r's axiom can then be formulated as :

\textbf{SO*} Let $x$, $y$, $x^{\prime}\in L$ be three orthogonal atoms then
there is $z\leq x\cup y$, such that the graph $\Gamma(x,z)$ is realizable in
$x\cup y\cup x^{\prime}$.

There is a way to construct the graph $\Gamma$ which will make \textbf{SO}*
obviously stronger than the original \textbf{SO}. To do this simply add to
$\Gamma$ the rays (and orthogonality relations) which force the relation
$\mathcal{H}(z;x,y)\ \bot\ z$. In the notations of section 2.4, this means
adding rays $u,v,s,t$ and also the rays which, in the space $x\cup y\cup
x^{\prime}$, are orthogonal to the planes $x\cup u$, $z\cup u$, $y\cup v$,
$x\cup s$, $z\cup v$, $u\cup t$ . But this is cheating, all it shows is that
there is a finite graph that forces Sol\`{e}r's axiom simultaneously with
uncertainty. In order to make the axiom more acceptable one has to solve

\begin{problem}
Find the minimal $\Gamma$ that forces logical indeterminacy and that allows
the proof of Sol\`{e}r's theorem (and even, perhaps, improves it to include
finite dimensional cases).
\end{problem}

Another possible candidate -analogous to Savage's axiom on the existence of
arbitrarily fine partitions- is the following:

\textbf{SO** }Let $z\in L$; then the Wondergraph $\Omega(z)$ is realizable in
any three dimensional subspace of $L$ that includes $z$.

The restriction of the graphs we have used to those realizable in
$\mathbb{R}^{3}$ is not essential. It may very well be that a more natural
candidate for our $\Gamma$ or $\Omega$ exists, e.g., in $\mathbb{C}^{4}$.

\section{Probability: Range and Classical Limit}

We turn now to the explanatory power of our analysis. The "logic of partial
belief" provides straightforward probabilistic, or even combinatorial
derivations of a variety of phenomena for which alternative approaches require
complicated ad-hoc dynamical explanations. We shall consider two central
examples: the first is the EPR paradox and the violation of Bell inequality,
and the second is the measurement problem. In particular, we shall discuss the
way macroscopic objects can be handled in this framework.

\subsection{Bell Inequalities}

The phenomenological difference between classical and quantum probability is
most dramatic when quantum correlations associated with entangled states are
concerned. Let us recall what the classical probabilistic analysis of the
situation is: A pair of objects is sent from the source, one in Alice's
direction, one in Bob's direction. Alice can perform either one of two
measurements on her object; she can decide to detect the event $x_{1}$ or its
absence (which means detecting the event $x_{1}^{\bot}$). Alternatively, she
can decide to check the event $x_{2}$ or $x_{2}^{\bot}$. So each of these
measurements has two possible outcomes. Similarly, Bob can test for $y_{1}$ or
use a different test to detect $y_{2}$. Assuming \emph{nothing} about the
physics of the situation, and just considering the outcomes we get the
following possible logical combinations expressed in the truth table:%

\[%
\begin{array}
[c]{cccccccc}%
x_{1} & x_{2} & y_{1} & y_{2} & x_{1}\cap y_{1} & x_{1}\cap y_{2} & x_{2}\cap
y_{1} & x_{2}\cap y_{2}\\
0 & 0 & 0 & 0 & 0 & 0 & 0 & 0\\
... & ... & ... & ... & ... & ... & ... & ...\\
1 & 1 & 0 & 1 & 0 & 1 & 0 & 1\\
... & ... & ... & ... & ... & ... & ... & ...\\
1 & 1 & 1 & 1 & 1 & 1 & 1 & 1
\end{array}
\]

It is the truth table of four propositional variables $x_{1},x_{2},y_{1}%
,y_{2}$ and four (out of the six) pair conjunctions, so it has $16$ rows,
three of them shown explicitly. Each row represents a possible state of
affairs regarding the possible outcomes where $1$ indicates that the event
occurs. Now, suppose that we were to bet on the outcomes. There are, of course
many ways to do this, but they all have to conform with the canons of
rationality. The only constraint here is that each one of the $16$
possibilities will be assigned a non-negative probability, and the sum of
these probabilities be $1$. To give this fact a geometric interpretation
consider each one of the $16$ rows in the truth table as a vector in an $8$
dimensional real space, then the vector of probabilities (writing $x_{i}y_{j}$
for $x_{i}\cap y_{j}$)%

\[
\mathbf{P}=(P(x_{1}),P(x_{2}),P(y_{1}),P(y_{2}),P(x_{1}y_{1}),P(x_{1}%
y_{2}),P(x_{2}y_{1}),P(x_{2}y_{2}))
\]
lies in the convex hull of these $16$ vectors, which is a correlation polytope
in $\mathbb{R}^{8}$ with the $16$ truth values as vertices shown schematically
in Fig(\ref{3}).%
\begin{figure}
[ptb]
\begin{center}
\includegraphics[
natheight=2.646300in,
natwidth=3.480000in,
height=2.6887in,
width=3.5267in
]%
{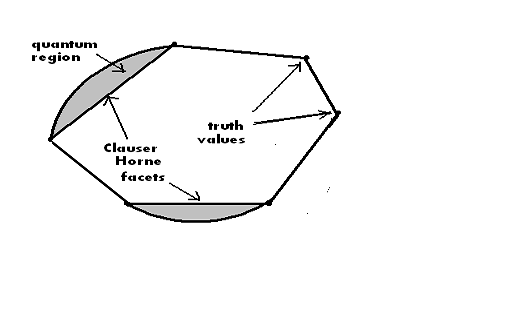}%
\caption{Classical and quantum correlations}%
\end{center}
\end{figure}

The facets of the polytope, are given by linear inequalities in the
probabilities, in this case the non-trivial inequalities have the form%

\begin{equation}
-1\leq P(x_{1}y_{1})+P(x_{1}y_{2})+P(x_{2}y_{2})-P(x_{2}y_{1})-P(x_{1}%
)-P(y_{2})\leq0
\end{equation}
They are called Clauser-Horne inequalities\footnote{The inequalities were
derived in \cite{34}. The sufficiency of the inequalities is due to Fine
\cite{35}. The polyhedral structure, its relation to logic, and its
generalizations are discussed in \cite{22, 36}.}, they are among what is
generally known as Bell inequalities. Remarkably, in the mid 19-century George
Boole considered the most general form of the constraints on the values of
probabilities of events that can be derived from the logical relations among
them. He proved that these constraints have the form of linear inequalities in
the probabilities. Paraphrasing Kant he called such constraints
\emph{Conditions of Possible Experience}\footnote{In \cite{37}, see also
\cite{8}. The parody of Kant is intended, I think. In his classic \emph{The
Laws of Thought} Boole writes: "Now what has been said,..., is equally
applicable to many other of the debated points in philosophy; such, for
instance, as the external reality of space and time. We have no warrant for
resolving these into mere forms of the understanding, though they
unquestionably determine the \emph{present} sphere of our\textrm{\ }knowledge"
(\cite{38}, page 418, my emphasis). So, in the end the joke is on Boole.\ }.

So far we have been concentrating on the classical picture. What is the
quantum mechanical analysis? Again, we shall make no physical assumptions
beyond those which are given by the axioms of the event structure. With the
two particles we associate a Hilbert space of the form $H\otimes H$, where in
case the objects are spin-%
${\frac12}$
particles, $\dim H=2$. The relevant lattice is thus $L=L(H\otimes H)$. The
element of $L$ corresponding to the event $x_{1}$ is a two dimensional
subspace of the form $a_{1}\otimes1$ where $a_{1}\in L(H)$ and $1$ is the unit
in $L(H)$. Similarly, the event corresponding to the outcome $y_{1}$ on Bob's
side is $1\otimes b_{1}$, and likewise for the other cases. The event
corresponding to the measurement of $x_{1}$ on Alice's side and $y_{1}$ on
Bob's side is just the intersection:%

\[
(a_{1}\otimes1)\cap(1\otimes b_{1})=a_{1}\otimes b_{1}%
\]
Note also that $a_{i}\otimes1$, and $1\otimes b_{j}$ are compatible. Now, to
the eight outcomes\ %

\[%
\begin{array}
[c]{cccccccc}%
a_{1}\otimes1, & a_{2}\otimes1, & 1\otimes b_{1}, & 1\otimes b_{2}, &
a_{1}\otimes b_{1}, & a_{1}\otimes b_{2}, & a_{2}\otimes b_{1}, & a_{2}\otimes
b_{2},
\end{array}
\]
correspond an $8$ dimensional vectors of probability values%

\[
\mathbf{P}=(P(a_{1}\otimes1),P(a_{2}\otimes1),P(1\otimes b_{1}),P(1\otimes
b_{2}),P(a_{1}\otimes b_{1}),P(a_{1}\otimes b_{2}),P(a_{2}\otimes
b_{1}),P(a_{2}\otimes b_{2})),
\]
where $P$ is any probability assignment to the elements of $L(H\otimes H)$.
When we vary $P$ and the subspaces $a_{i}$, $b_{j}$, we see that the quantum
range is larger than the classical one, and some points lie outside the
classical polytope (Fig \ref{3}), that is, they violate one of the facet
inequalities of Clauser and Horne.

From the point of view developed so far this consequence is natural and
follows from the event structure of quantum mechanics via Gleason's theorem.
We also know from corollary 10 that a violation of a Clauser-Horne inequality
can already be depicted in a finite gamble (an explicit construction can be
found in \cite{15}). Altogether, in our approach there is no problem with
locality and the analysis remains intact no matter what the kinematic or the
dynamic situation is; the violation of the inequality is a purely
probabilistic effect. Notice that we are just using the quantum event space
notion of intersection between (compatible) outcomes: $(a_{1}\otimes
1)\cap(1\otimes b_{1})=a_{1}\otimes b_{1}$, as we have used the intersection
in the classical event space. The derivation of Clauser-Horne inequalities,
indeed of many of Boole's conditions, is blocked since it is based on the
Boolean view of probabilities as weighted averages of truth values. This, in
turn, involves the metaphysical assumption that there is, simultaneously, a
matter of fact concerning the truth values of incompatible propositions such
as $x_{1}=a_{1}\otimes1$ and $x_{2}=a_{2}\otimes1$.

Recall that in section 2.2 we restricted "matters of fact" to include only
observable records. Our notion of \textquotedblleft fact\textquotedblright\ is
analytically related to that of \textquotedblleft event\textquotedblright\ in
the sense that a bet can be placed on $x_{1}$ only if its occurrence, or
failure to occur, can be unambiguously recorded. However, this leaves open a
metaphysical question: Given that $x_{1}$ occurred, what is the status of
$x_{2}$ for which no observable record can exist? Our axioms are not designed
to rule out the possibility that $x_{2}$ \emph{has} a truth value which we do
not know. Initially our approach was agnostic with respect to facts which
leave no trace. However, as the above analysis shows, assigning truth values
to $x_{2}$ and $x_{1}$ simultaneously is untenable. In other words, it is
prohibited by the axioms \emph{a posteriori}.\footnote{This also follows from
the logical indeterminacy principle (theorem 3) or the (weaker) Kochen and
Specker's theorem \cite{33}.} I believe that Bohr deserves the credit for this
insight, although his arguments fall short of establishing it.

We should also recall that there are alternatives to quantum mechanics in
which the violations of the Clauser-Horne inequalities have non-local
dynamical origins. However, from our perspective the commotion about locality
can only come from one who sincerely believes that Boole's conditions are
really conditions of possible experience. Since these conditions are just
properties of the classical intersection of events, their violation must
indicate that something is not kosher with the \emph{measurements}, that is,
the choice of a measurement on one side may be correlated with the outcome on
the other. But if one accepts that one is simply dealing with a different
notion of probability, then all space-time considerations become irrelevant.

\subsection{The BIG measurement problem, the small one, and the classical
limit.}

There are two \textquotedblleft measurement problems\textquotedblright\ The
BIG problem, which is illusory, and the small problem which is real and
concerns the quantum mechanics of macroscopic systems. The BIG problem
concerns those who believe that the quantum state is a real physical state
which obeys Schr\"{o}dinger's equation in all circumstances. In this picture a
physical state in which my desk is in a superposition of being in Chicago and
in Jerusalem is a real possibility; and similarly a superposed alive-dead cat.
In fact the linearity of Schr\"{o}dinger's equation implies that (decoherence
notwithstanding) it is easy to produce states of macroscopic objects in
superposition- which seems to contradict our experience, and sometimes, as in
the cat case, does not even make much sense.

In our scheme quantum states are just assignments of probabilities to possible
events, that is, possible measurement outcomes. This means that the updating
of the probabilities during a measurement follows the Von Neumann-L\"{u}ders
projection postulate and not Schr\"{o}dinger's dynamics. Indeed, the
projection postulate is just the formula for conditional probability that
follows from Gleason's theorem. So the BIG measurement problem does not arise.
In particular, the cat in the Schr\"{o}dinger thought experiment is not
superposed, but is rather cast in the unlikely role of a particle spin
detector. Schr\"{o}dinger's equation governs the dynamics between
measurements; it dictates the way probability assignments should change over
time in the absence of a measurement. The general shape of the
Schr\"{o}dinger's equation is not a mystery either; the unitarity of the
dynamics follows from the structure of $L(\mathbb{H})$ via a theorem of Wigner
\cite{39}, in its lattice theoretic form \cite{40}. However, these remarks do
not completely eliminate the measurement problem because in our scheme quantum
mechanics is also applicable to macroscopic objects.

So suppose that $x$ is one of the rays in the cat's Hilbert space
corresponding to a living cat. Let $y$ be one of the atoms corresponding to a
dead cat so that $x\bot y$. By Sol\`{e}r's axiom there is an atom $z\leq x\cup
y$ which bisects the angle between $x$ and $y$. Does this mean that we are
back with the BIG measurement problem? The answer is `No'; remember that $z$
is not a state of the system, it is a possible measurement outcome. It is a
mistake to think that by merely following Schr\"{o}dinger's experiment we are
"observing" the event $z$, or something like it. Obviously we are not, we
either see an $x$-like event, a live cat, or a $y$-like dead cat event. In
order to "see" $z$ we have to devise and perform a measurement such that $z$
is one of its eigenspaces. For reasons that will be explained below, with all
probability this is impossible.

But even agreeing that \emph{performing} such a measurement is impossible, we
can surely think about operators for which $z$ is an eigenspace, say the
projection on $z$. So let us imagine what one will see when one performs this
measurement; what does the event $z$ look like? Presumably, the imagined
measuring device is a huge piece of very complicated equipment, because in all
likelihood the measurement of the projection on $z$ involves manipulating
individual cat particles. In the end, however, there is a dial with two
possible readings $0$ and $1$, and $z$ is just the event that the dial reads
$1$. By L\"{u}ders' rule the state of the cat after the measurement -assuming
that $z$ was the outcome- is the projection on $z$. The quantum state is not a
physical object, it is a representation of our state of knowledge, or belief.
The projection on $z$ represents an extremely complex assignment of
probabilities to all possible events in a Hilbert space of $\backsim10^{25}$
particles, an intractable business. One thing is clear, though, there is
complete uncertainty about the cat being dead or alive $P(x)=P(y)=\frac{1}{2}%
$, and of course $P(x\cup y)=1$.

Ignorance aside, is it not the case that now, after the measurement, there is
a matter of fact about the cat being dead or alive? Well, No! As in all such
circumstances we cannot say that there is a fact regarding this matter. It is
impossible in principle to obtain a record concerning the cat being alive or
dead simultaneously with the $z$-measurement. There is no fundamental
difference between the present case and EPR, meaning that we cannot
consistently maintain that the proposition \textquotedblleft the cat is
alive\textquotedblright\ has a truth value. But the devil is in the details;
there is no way to tell from our completely schematic description what is
going on in the laboratory. Consequently, there is no way to tell what is the
biological state of the cat. It is only after we have mastered the details of
the measuring process that we can understand the exact sense in which no
record of $x$ or of $y$ is obtainable.

\subsection{The Weak Entanglement Conjecture}

The small measurement problem is the question why we do not routinely observe
events like $z$ for macroscopic objects. More precisely why is it hard to
observe macroscopic entanglement, and what are the conditions in which it
might be possible? One answer which is certainly valid is decoherence- meaning
that it is extremely hard to isolate large pieces of matter and equipment from
environmental noise. Decoherence is a dynamical process and its exact
character depends on the physics of the situation. I would like to point a
possible more fundamental, purely combinatorial reason which is an outcome of
the probabilistic structure: \emph{The entanglement of an average ray in a
multiparticle Hilbert space is very weak}. To make this intuition precise we
have to quantify entanglement, and define what we mean by \textquotedblleft
average ray\textquotedblright.

To make the discussion simpler we shall concentrate on qbits. So our Hilbert
space is composed of $n$ copies of the two dimensional complex Hilbert space
$\mathbb{H}_{n}\mathbb{=C}^{2}\otimes\mathbb{C}^{2}\otimes...\otimes
\mathbb{C}^{2}$, and $\dim\mathbb{H}_{n}=2^{n}$.\ An atom $s\in L(\mathbb{H}%
_{n}\mathbb{)}$ is called \emph{separable} if it has the form $s=x_{1}\otimes
x_{2}\otimes...\otimes x_{n}$ with $x_{i}\in L(\mathbb{C}^{2})$, otherwise an
atom is called \emph{entangled}. Also, we shall call the projections on
separable (entangled) rays, separable (entangled, respectively) pure states.
We keep the letter $s$ to designate separable atoms, and denote by $S\subset
L(\mathbb{H}_{n}\mathbb{)}$ the set of all separable atoms. As usual if $x\in
L(\mathbb{H}_{n}\mathbb{)}$ is a ray (atom), we shall denote by
$\overrightarrow{x}$ a \emph{unit} vector along it.

Now, suppose that we want to observe an entangled atom $x$. More precisely, we
want to obtain a positive proof that it is indeed entangled. To do this we
have to design a measurement that will distinguish the ray $x$ from all the
separable atoms $s\in S$. A Hermitian operator that does this always exists,
and will be called an \emph{entanglement witness for }$x$, or in short, a
witness. The normalization of witnesses is a matter of convention and for our
purpose we shall use the following:

\begin{definition}
An Hermitian operator $W$ on $\mathbb{H}_{n}\mathbb{=C}^{2}\otimes
\mathbb{C}^{2}\otimes...\otimes\mathbb{C}^{2}$ ($n$ copies) is called an
entanglement witness if it satisfies
\[
\sup\{\left\vert \left\langle \overrightarrow{s},W\overrightarrow
{s}\right\rangle \right\vert \ ;\ s\in\nolinebreak S\}=\nolinebreak1
\]
while
\[
\left\Vert W\right\Vert =\sup\{\left\vert \left\langle \overrightarrow
{x},W\overrightarrow{x}\right\rangle \right\vert \ ;\ x\in L(\mathbb{H}%
_{n}\mathbb{)\}}>1.
\]

\end{definition}

So a witness is an observable whose expectation on every separable state is
bounded between $-1$ and $1$, while it has an eigenvalue that is larger than
$1$ in absolute value. Any one-dimensional eigenspace $x$ corresponding to
this eigenvalue is obviously entangled. Denote by $\mathcal{W}_{n}$ the set of
all entanglement witnesses on $\mathbb{H}_{n}$. One way to estimate \emph{how
much} a given $x\in L(\mathbb{H}_{n}\mathbb{)}$ is entangled is to calculate%

\begin{equation}
\mathcal{E}(x)=\sup\{\left\vert \left\langle \overrightarrow{x}%
,W\overrightarrow{x}\right\rangle \right\vert \ ;\ W\in\mathcal{W}%
_{n}\mathcal{\}}%
\end{equation}
A witness $W$ at which the value $\mathcal{E}(x)$ obtains is the best witness
for the entanglement of $x$. If we allow that every measurement involves
errors then the larger $\mathcal{E}(x)$ is, the more likely we are to actually
observe it. The good news is that \emph{there are rays }$x\in L(\mathbb{H}%
_{n})$\emph{\ such that }$\mathcal{E}(x)=\sqrt{2^{n}}$. These correspond to
the the maximally entangled states, the so-called generalized GHZ
states\footnote{See Mermin \cite{41}. The witnesses that provide the maxumum
value have a close relation to the facets of the correlation polytope for this
case, see \cite{42, 43}.}. However, it seems that such rays become more and
more rare as $n$ increases. To formulate this intuition precisely, let
$\mu_{n}$ be the \emph{normalized} uniform (Lebesgue) measure on the unit
sphere of $\mathbb{H}_{n}$. Then we

\begin{conjecture}
There is a universal constant $C>0$ such that
\begin{equation}
\mu_{n}\left\{  \overrightarrow{x}\ ;\ \mathcal{E}(x)>C\sqrt{n\log n}\right\}
\rightarrow0\quad as\ n\rightarrow\infty
\end{equation}

\end{conjecture}

A similar result has been established for a large family of witnesses that for
each $n$ contains $2^{2^{n}}$ witnesses, and which include those that give the
best estimation for the GHZ states \cite{44}; hence the conjecture.

I think the conjecture, if true, concerns our ability to observe macroscopic
entanglement. There are two types of macroscopic or mesoscopic rays whose
entanglement might be witnessed, and the conjecture concerns the second case:

\textbf{1}. There may be relatively rare cases in which the entanglement
witness happens to be a thermodynamic observable, that is, an observable whose
measurement does not require manipulation of individual particles but only the
observation of some global property of the system. There are some indications
that this may be the case for some spin chains and lattices \cite{45}.

\textbf{2}. Cases of very strong entanglement, like GHZ, which do require many
manipulations of individual particles to be observed; however, the value of
$\mathcal{E}(x)$ is large enough to give significant results that rise above
the measurement errors. If we assume that the measurement errors are
independent, then the total expected error grows exponentially with the number
of particles that are manipulated. So, in general, one expects that only $x$'s
for which $\mathcal{E}(x)$ is exponential in the number of manipulated
particles could yield a significant outcome. The conjecture proposes that the
proportion of such $x$ 's is low.

To sum up: the answer to the question "why don't I see chairs in
superposition" is twofold, decoherence surely, but even if we could turn it
off, there is the combinatorial possibility that "seeing" something like this
is nearly impossible. All this, luckily, does not prevent the existence of
exotic macroscopic superpositions that can be recorded.

\section{Measurements}

In this paper, all we have discussed is the Hilbert space formalism. I have
argued that it is a new kind of probability theory that is quite devoid of
physical content, save perhaps the indeterminacy principle which is built into
axiom \textbf{H4}. Within this formal context there is no explication of what
a measurement is, only the identification of "observables" as Hermitian
operators. In this respect the Hilbert space formalism is really just a syntax
which represents the set of all possible outcomes, of all possible
measurements. It is analogous to the mathematical concept of a probability
space, in which certain subsets are identified as events. However, the
mathematical theory of probability itself does not tell us the nature of the
connection between these formal creatures and real events in the world.

But even before a connection is made between the formal and physical sense of
measurement I think there is an interesting philosophical problem here. Our
formalism seems to be consistent: there is a \emph{possible} world where
measurements and their outcomes behave in the way described above. This would
not have been a serious problem if the classical theory of probability were
not conceived as \emph{a priori} in some sense. But the theory of probability
is a part of what we take as our theory of inference, hence the term
`\emph{logic} of partial belief'. As such it is also a ground for the
formation of rational expectations. Therefore, the fact that there is a
consistent alternative poses a problem similar to the problem that
non-Euclidean geometry raised even before general relativity. What should we
make of a world in which Boole's conditions of possible experience are
violated for no reason other than the structure of probabilities described here?

What is real in the quantum world? Firstly, there are objects- particles about
which the theory speaks- which are identified by a set of parameters that
involve no uncertainty, and can be recorded in all circumstances and thus
persist through time and context \cite{46}. Among them are the rest mass,
electric charge, baryonic number, etc. The other part of quantum reality
consists of events, that is, recordings of measurements in a very broad sense
of the word. Now, one has to distinguish between measurements on the one hand
and interactions between material objects on the other. The latter are best
described in the Heisenberg picture: There is a time dependent interaction
Hamiltonian $H(t)$ which, like any other observable, defines at every moment
$t$ a set of \emph{possible} outcomes, one of the outcomes would obtain
\emph{if} $H(t)$ were measured. If, in addition, we have formed a belief about
the state of the system at time $t=0$ (as a result of a previous measurement,
say) we automatically have a probability distribution over the set of all
possible outcomes of all possible measurements at each $t$. So each
interaction constrains the set of possible outcomes in a certain specific way,
and the question which interactions can actually be executed is an empirical
question, to be tested by observing the outcomes and their distributions.
Measurements are \emph{not} interactions in this sense; although in the broad
description of an experiment there is usually an interaction leading to the measurement.

It is impossible to give a precise definition of all the physical processes
that deserve the name \emph{measurement}; just as it is not possible to define
the term \emph{event} to which the theory of probability can be applied. Even
a non-contextual definition of a singular concrete measurement is hard to
provide; in this sense measurement outcomes are events "under a description",
as philosophers say. Broadly speaking, a measurement is a process in which a
material system $M$, prepared in a specific way, records some aspect of
another system $S$, a recording that effects a permanent change in $M$, or at
least one that lasts long enough. The outcomes to which we have referred
throughout the paper are such recordings. Probably the best way to describe
measurements is in informational terms. The information recorded by a
measurement is \emph{systematic} in the sense that a repeated conjunction of
$M$ and $S$ yields the same set of results, and the frequency distribution
over the set of results stabilizes in the long run. Of the same importance is
the information that is lost during a measurement, the outcome that we could
have obtained if any other measurement $M^{\prime}$ were performed instead of
$M$ \cite{47}.

This description is broad enough to include the change that photons imprint on
the receptors of the retina; it also includes the change caused by a proton
hitting a rock on the dark side of the moon. There is nothing specifically
human about measurements, nor does $M$ have to be associated with a
macroscopic system. What constitutes a "measuring device" cannot be determined
beyond this broad description. However, there is a structure to the set of
events. Not only does each and every type of measurement yield a systematic
outcome; but also the set of all possible outcomes of all measurements
-including those that have been realized by an actual recording- hang together
tightly in the structure of $L(\mathbb{H})$. This is the quantum mechanical
structure of reality.

\bigskip

\textbf{Acknowledgements:} This paper is the outcome of a three lecture series
that I gave at the Patrick Suppes Center for the Interdisciplinary Study of
Science and Technology, Stanford University. I would like to thank Patrick
Suppes, Michael Friedman and Thomas Ryckman for their generous hospitality and
the lively discussions. I also want to thank William Demopoulos and Ehud
Hrushovski for many conversations on questions of philosophy, logic, and
mathematics, and Jeremy Butterfield for his comments and suggstions. The
research leading to this paper is supported by the Israel Science Foundation
grant number 879/02.

\end{document}